\documentclass[%
 reprint,
 amsmath,amssymb,
 aps,
]{revtex4-1}

\usepackage{graphicx}% Include figure files
\usepackage{dcolumn}% Align table columns on decimal point
\usepackage{bm}% bold math
\usepackage{xcolor}
\usepackage{xspace}
\usepackage[normalem]{ulem}
\usepackage{xfrac}
\usepackage{fontawesome}

\usepackage[babel]{csquotes}
\MakeOuterQuote{"}

\providecommand{\sorthelp}[1]{}
\makeatletter
\let\jnl@style=\rmfamily
\def\ref@jnl#1{{\jnl@style#1}}%
\newcommand\aap{\ref@jnl{A\&A}}%
\makeatother

\usepackage[bookmarks]{hyperref}% add hypertext capabilities
\hypersetup{
    colorlinks,
    linkcolor={red!50!black},
    citecolor={blue!50!black},
    urlcolor={blue!80!black}
}

\newcommand{\planck}{{\it Planck}\xspace}

\newcommand{\hubbleunit}{km/s/Mpc\xspace}

\newcommand{\comment}[1]{{}}

\newcommand{\be}{\begin{equation}}
\newcommand{\ee}{\end{equation}}
\newcommand{\bea}{\begin{eqnarray}}
\newcommand{\eea}{\end{eqnarray}}

\newcommand{\lcdm}{$\Lambda$CDM\xspace}
\newcommand{\LCDM}{$\Lambda$CDM\xspace}

\newcommand{\fracdelta}[1]{\delta #1 / #1}

\def\oz#1{{}}

\defcitealias{aylor2019}{A19}
\defcitealias{riess2019}{R19}

\begin{document}

\preprint{APS/123-QED}

\title[The Hubble Hunter's Guide]{The Hubble Hunter's Guide}% Force line breaks with \\
\thanks{with apologies to \citet{gunion1989}}

\author{L. Knox}
\altaffiliation{Department of Physics, University of California, One Shields Avenue, Davis, CA 95616, USA}

\author{M. Millea}
\altaffiliation{Berkeley Center for Cosmological Physics and Department of Physics, University of California,
Berkeley, CA 94720, USA}

\date{\today}% It is always \today, today,

\begin{abstract}
Measurements of the Hubble constant, and more generally measurements of the expansion rate and distances over the interval $0 < z < 1$, appear to be inconsistent with the predictions of the standard cosmological model (\lcdm) given observations of cosmic microwave background temperature and polarization anisotropies. Here we consider a variety of types of departures from \lcdm that could, in principle, restore concordance among these datasets, and we explain why we find almost all of them unlikely to be successful. We single out the set of solutions that increase the expansion rate in the decade of scale factor expansion just prior to recombination as the least unlikely. These solutions are themselves tightly constrained by their impact on photon diffusion and on the gravitational driving of acoustic oscillations of the modes that begin oscillating during this epoch -- modes that project on to angular scales that are very well measured. We point out that a general feature of such solutions is a residual  to fits to \LCDM, like the one observed in Planck power spectra. This residual drives the modestly significant inferences of angular-scale dependence to the matter density and anomalously high lensing power, puzzling aspects of a data set that is otherwise extremely well fit by \lcdm.
\end{abstract}

\keywords{Suggested keywords}%Use showkeys class option if keyword
\maketitle

%new cite requests:
%[x]"I'm writing to bring your attention to a paper written myself and Stephon Alexander, https://arxiv.org/abs/1904.08912 , which predates your reference to Lin et al [20] by over a month."
% [x]Desmond, Jain, Sakstein
%[x] Add L. Page to acknowledgments
%[] Marius funding source to acknowledgments?
%[x]Poulin email
%[x]In the light relics part we should say something about interacting neutrinos. Somewhere we should make reference to the interacting DM-DR model and paper with Buen-Abad, Lesgourgues and Schmaltz. The latter should perhaps be in the introduction.

\section{Introduction}

Estimates of the Hubble constant from a distance ladder approach are  generally higher than those derived from cosmic microwave background (CMB) data, assuming the standard ``\lcdm'' cosmological model \citep{planck2016-l06}. The SH$_0$ES team calibrates a supernova sample with Cepheids and finds \mbox{$H_0 = 74.03 \pm 1.42$ \hubbleunit} \cite[hereafter \citetalias{riess2019}]{riess2019}.
Compared with the value inferred from Planck CMB temperature and polarization power spectra plus CMB lensing, assuming \LCDM, \mbox{$H_0 = 67.27 \pm 0.60$ \hubbleunit}, there is a $4.4\,\sigma$ discrepancy. The most recent result from strong-lensing time delays, from the H$_0$LiCoW team \citep{chen2019,wong2019}, assuming the standard cosmological model and a prior of $\Omega_{\rm m} \in [0.05,0.5]$, of $H_0 = 76.8 \pm 2.5$ \hubbleunit\  is consistent with \citetalias{riess2019} and discrepant with the \lcdm Planck value at $3.1\,\sigma$. The Carnegie-Chicago Hubble Project have used their own Hubble flow set of supernovae that they have calibrated with the tip of the red giant branch method. They find \mbox{$H_0 = 69.8 \pm 0.8 ({\rm stat}) \pm 1.7 ({\rm sys})$ \hubbleunit} \cite{freedman2019}, which at the $2\,\sigma$ level is consistent with all of these results. 

\citet{bernal2016} showed that in addition to a discrepancy in the Hubble constant, there is a discrepancy in the comoving sound horizon at the end of the baryon drag epoch, $r_{\rm s}^{\rm drag}$. They used Cepheid-calibrated supernovae \cite{riess2016,betoule2014} to infer distances out to redshifts with precise measurements of the baryon acoustic oscillation feature \cite{beutler2011,ross2015,kazin2014}. With these distances, they could convert the BAO angles to inferences of $r_{\rm s}^{\rm drag}$. Using supernova data to control the shape of $D(z)$ they obtained relatively model-independent inferences of this empirically-determined sound horizon and showed that it is lower than the \lcdm Planck-determined sound horizon by 7\%, amounting to a $2.6\,\sigma$ difference. 

\citet[hereafter \citetalias{aylor2019}]{aylor2019} repeated this analysis with updated data, and found the sound horizon tension to be robust to choice of CMB dataset, and thereby argued against systematic errors in CMB data as a source of the discrepancy~\footnote{Similar arguments have been made with the inverse distance ladder approach \citep{percival2010, heavens2014, aubourg2015, cuesta2015, bernal2016, macaulay2019, verde2017, lemos2019, joudaki2018, feeney2018}.}. \citetalias{aylor2019} argued that the sound horizon tension implies that any cosmological solution to the discrepancy between distance ladder and CMB measurements is likely to include changes to the cosmological model in the two decades of scale factor evolution prior to recombination. 

Many attempts at solutions do indeed take this route. These include the extension of additional light relics \cite[e.g.]{planck2013-p11} and an extension to include what is traditionally called "early dark energy." For examples of the latter see \cite{karwal2016, evslin2018, poulin2018, agrawal2019,rossi2019,alexander2019,lin2019}. The authors of \cite{agrawal2019} find $66.7 < H_0 < 70.6$~\hubbleunit\ (95\% confidence region) from CMB + BAO + uncalibrated supernova data. Other pre-recombination efforts include \cite{lin2019a} who propose a modified gravity solution. 

The well-motivated extension to light relics has become more tightly constrained as the CMB data improve. The current constraints from the combination of BAO and \planck temperature, polarization, and lensing power spectra, is $H_0 = 67.3 \pm 1.1$\,\hubbleunit \cite{planck2016-l01}. Modifications in the light relic sector are being explored in order to circumvent these bounds. These extensions include the introduction of strong scattering interactions between the neutrinos or between other additional light relics \cite{cyr-racine2014a,lancaster2017,kreisch2019} or between the dark matter and the tightly coupled gauge bosons of an associated hidden sector gauge field \cite{buen-abad2018a}. \citet{kreisch2019} find that by including interactions in the light relics sector (including the neutrinos) and allowing $N_{\rm eff}$ to be a free parameter, a combination of Planck data, BAO data, and the Hubble constant measurement from \cite{riess2016} yields $H_0 = 72.3 \pm 1.4$ \hubbleunit~\footnote{The reduction in tension here is hard to infer from this result because of the inclusion of the prior on $H_0$ in the reported $H_0$ determination}.  

Others have been motivated to pursue late-time solutions (model changes that are not important prior to recombination), despite the conclusions in \citetalias{aylor2019}, citing the challenges faced by the pre-recombination solutions. For example \cite{divalentino2017, divalentino2016a, joudaki2018} use an extended parameter space and point out that an interacting phantom-like dark energy  with equation of state $w_{\rm DE}<-1$ can reduce the tension in $H_{0}$ measurements. More recent attempts at late-time solutions include \cite{keeley2019} and \cite{raveri2019}. 

In this paper we revisit the claim of \citetalias{aylor2019} about where in redshift the departures from \lcdm need to be important. We do so by thinking as broadly as we can about possible solutions, and the measurements that constrain them. We qualitatively assess the challenges they face and their likelihood of successful implementation in a specific model. We intend our analysis to be a guide to further theoretical exploration of possible cosmological solutions to the $H_0$ discrepancy. We also hope to provide all readers with an appreciation of the significant challenges confronted by any model builder looking for cosmological solutions.

In considering the challenges faced by the pre-recombination alteration solutions we have been led to an interesting conclusion: these solutions generically lead to features in the CMB power spectra that we may already be seeing in the {\it Planck} data. Oscillatory residuals to the \lcdm fit to the Planck temperature power spectrum are responsible for an anomalously high preference for additional lensing power parameterized by $A_{\rm L}$. \citet{planck2016-l06} find that from the Planck 2018 TT+TE+EE+lowE data that there is a preference for excess lensing power of between $2\,\sigma$ and $3\,\sigma$ depending on which likelihood is used. These oscillatory residuals are also partially responsible for some mild tension ($2.3\sigma$) between estimates of the matter density inferred from different $\ell$ ranges \cite{planck2016-LI,addison2015,planck2016-LI}.

In Section II we review how $H_0$ is determined from CMB data under the assumption of \LCDM. In Section III we review constraints in the $r_{\rm s}^{\rm drag}-H_0$ plane from SH$_0$ES, BOSS BAO \cite{alam2017}, Pantheon Supernovae \cite{scolnic2018} and CMB data \cite{planck2016-l01}. The constraints in this plane help us to understand the challenge of reconciling these 4 data sets, and why we are driven toward solutions that reduce the sound horizon.

In section IV we revisit the question of solutions that do not change cosmology prior to recombination, discuss two such solutions in the literature \cite{joudaki2018,keeley2019,raveri2019}, and introduce and discuss a couple of exotic scenarios. In section V we consider solutions that {\em do} make changes to the \lcdm model prior to recombination. We group these into four categories: sound speed reduction, ``confusion sowing,'' high temperature recombination, and increased $H(z)$. We consider three different ways of achieving this time reduction: high-temperature recombination, faster-than-adiabatic photon cooling, and additional components to increase the expansion rate at a given temperature. 

For all of these classes of models we present the challenges to successful implementation. Reducing the sound horizon pushes us toward changes in either recombination, or in the ingredients of the model in the decade of scale factor evolution immediately prior to recombination. The observed CMB spectra are highly sensitive to the process of recombination, as photon diffusion has a large impact on the spectra and the damping tail has been measured very precisely. The observed CMB spectra are also highly sensitive to the acoustic dynamics of modes that begin oscillating in that final decade of scale factor evolution prior to recombination. In short, it is hard to reduce the sound horizon without creating additional consequences that disagree with observations.

%Should say more about how we explore additional components.
In assessing the observable consequences of additional contributions to the expansion rate near the epoch of recombination, we point out the important role played by the radiation-driving phenomenon \citep{hu1997}. In section VII we note that for data generated in a universe with additional components, we generically expect that an analysis of such data assuming \lcdm, would lead to angular scale-dependent inferences of the matter density. We then summarize the existing evidence for such scale dependence, before giving final remarks and conclusions. Code to produce the figures in this paper is available here: \href{https://github.com/marius311/hubblehunters/releases/tag/arxiv_v1}{\faGithub} \footnote{\url{https://github.com/marius311/hubblehunters/releases/tag/arxiv_v1}; We acknowledge use of \citet{cosmoslik}.}.

{
\renewcommand{\arraystretch}{1.4}
\begin{table*}[]
\begin{tabular}{l|l}
\toprule
Quantity & Description \\
\colrule
$z_{\rm drag}$, $z_\star$, $z_{\rm EQ}$ & Redshift of the baryon drag epoch, of CMB last-scattering, and of matter-radiation equality \\
$r_{\rm s}^{\rm drag}$, $r_{\rm s}^\star$, $r_{\rm s}^{\rm EQ}$ &  Comoving sound horizon at these three respective redshifts (see Eqn.~\ref{eqn:soundhorizonintegral}) \\
$r_{\rm d}^\star$ & Comoving diffusion length at CMB last-scattering \\
$\bar r_{\rm s}$, $\bar r_{\rm d}$ & Visibility-averaged sound horizon and diffusion length, respectively (see Appendix.~\ref{app:visrsrd}) \\
$D_{\!A}^\star$ & Angular diameter distance to last-scattering (see Eqn.~\ref{eq:DA})\\
$\theta_{\rm d}^\star$, $\theta_{\rm s}^\star$, $\theta_{\rm s}^{\rm EQ}$ &  Angular size on the last-scattering surface of $r_{\rm d}^\star$, $r_{\rm s}^\star$, and $r_{\rm s}^{\rm EQ}$, respectively \\
$\omega_x(\equiv\Omega_x h^2)$ & Cosmological density today of ingredient ``$x$'' in units of $\simeq 1.878\times10^{-26} \; \rm kg/m^3$  \\
\botrule
\end{tabular}
\caption{Summary of different symbols used in this work. Throughout, $r$ refers to a comoving length scale, $k$ to a comoving wave number, $\theta\equiv r/D_{\!A}^\star$ to an angular scale on the last-scattering surface, and $\ell$ to a multipole moment. Subscripts refer to the quantity being integrated, i.e. ``s'' for the sound horizon or ``d'' for the diffusion length. Super scripts (where applicable) refer to the limit of integration.}
\end{table*}
}

\section{Estimating $H_0$ from CMB Data Assuming \LCDM}

In practice, to determine $H_0$ from CMB data one calculates a Monte Carlo Markov Chain (MCMC) which involves evaluation of the likelihood of parameter values (and their associated spectra) at tens to hundreds of thousands of points in the parameter space, and then one uses this chain to infer the posterior density of  $H_0$, or any other cosmological parameter of interest. It is possible to perform this calculation, and get reliable results, without any thought about what is happening physically. Here our goal is to provide some physical insight into what makes it possible to constrain $H_0$ from CMB data, given the \lcdm model.

We can think of the estimation of $H_0$ from CMB data as proceeding in three steps: 1) determine the baryon density and matter density to allow for calculation of $r_{\rm s}^\star$, 2) infer $\theta_{\rm s}^\star$ from the spacing between the acoustic peaks to determine the comoving angular diameter distance to last scattering $D_{\!A}^\star = r_{\rm s}^\star/\theta_{\rm s}^\star$, 3) adjust the only remaining free density parameter in the model so that $D_{\!A}^\star = \int_{0}^{z_*} dz/H(z)$ gives this inferred distance. With this last step complete we now have $H(z)$ determined for all $z$, including $z\,{=}\,0$. We now describe these steps in more detail.

\subsection{Calibrating the Ruler}
\label{sec:calibrate_ruler}

The sound horizon at CMB last-scattering is
\be
\label{eqn:soundhorizonintegral}
r_{\rm s}^\star = \int_0^{t_\star} \frac{dt}{a(t)} \; c_s(t) = \int_{z_\star}^\infty \frac{dz}{H(z)} \; c_s(t),
\ee
where $t_\star$ and $z_\star$ are the time and redshift for which the optical depth to Thomson scattering reaches unity. This scale is closely related to $r_{\rm s}^{\rm drag}$, which is given by the same integrand but instead integrated until the end of the baryon drag epoch, which comes slightly later. It is $r_{\rm s}^{\rm drag}$ that is relevant for baryon acoustic oscillations and $r_{\rm s}^\star$ for the CMB power spectra. Despite their significant difference (the latter is about 2\% smaller in \lcdm) we expect a negligible amount of model dependence in this difference. Hence, for the remainder of the work, we implicitly assume that knowing one quantity allows us to determine the other in a way that is unlikely to be significantly affected by any possible cosmological solution to the tension.

In the \lcdm model, the cosmological parameter dependence of $t_\star$ and $z_\star$ is also small enough as to have a subdominant impact on the parameter dependence of $r_{\rm s}^\star$, and we neglect this dependence for the remainder of this subsection. In terms of determining $r_{\rm s}^\star$ from Eqn.~\eqref{eqn:soundhorizonintegral}, that leaves us with $c_s(z)$ and $H(z)$. The sound speed depends on the ratio of baryon density to photon density. In the standard cosmological model the radiation density is entirely determined by the highly precise determination of the temperature of the CMB \cite{fixsen1996,wright1994}; thus we can think of $c_s(z)$ as dependent on $\omega_{\rm b}$ alone. To determine $H(z)$ we need to know the mean densities. The remainder of the radiation density is determined by assumptions that end up determining neutrino density as a function of photon temperature. The only other densities that are free parameters are the matter density and the energy density associated with the cosmological constant. Of these two, only the matter density affects the sound horizon. 

Thus, to determine $r_{\rm s}^\star$, we only need to know the values of  $\omega_{\rm b}$ and $\omega_{\rm m}$. 
A review of the physics of CMB parameter estimation in \lcdm was recently given in Section 4 of \cite{planck2016-LI}. Here we briefly summarize this for just these two parameters.

One can estimate $\omega_{\rm m}$ from CMB power spectra due to its impact on the early Integrated Sachs-Wolfe (ISW) effect, the ``potential envelope'', and, the gravitational-lensing-induced smoothing of the spectra. The most precise determinations of $\omega_{\rm m}$ to date depend primarily on the potential envelope effect, so we focus on that here. As a given Fourier mode crosses the horizon, the resulting gravitational potential decay provides a near-resonant driver of the oscillation. The greater the ratio of matter to radiation at horizon crossing, the less the decay, and the lower the amplitude of the resulting oscillation. The ``potential envelope'' refers to the scale-dependent boosting of oscillation power, a boost that slowly plateaus to a peak value at angular scales smaller than $\theta_{\rm s}^{\rm EQ}$, the angular extent of the sound horizon at matter-radiation equality, projected from the last-scattering surface \footnote{We note though that the earlier work used the particle horizon at matter-radiation equality rather than the sound horizon at matter-radiation equality. Both the particle horizon and the sound horizon at matter-radiation equality are important scales for the radiation-driving effect, and the particle horizon is the important scale for the early ISW effect. The distinction is not too important in practice though, as, to a very good approximation they only differ by a factor ($\sqrt{3}$) that is independent of cosmological parameters, absent radical changes to the sound speed.}.

In the \lcdm model, $\theta_{\rm s}^{\rm EQ}$ depends primarily on $\rho_{\rm m}/\rho_{\rm rad}$ (with additional weak dependence on $\rho_{\rm b}/\rho_\gamma$ and $\Omega_{\rm m}$). Thus, from the impact of the potential envelope on the spectra, we can infer $\theta_{\rm s}^{\rm EQ}$ and then $\rho_{\rm m}/\rho_{\rm rad}$. Given that $\rho_{\rm rad}$ is completely determined in \lcdm by the temperature of the CMB we can thus infer the matter density.

For determining $\omega_{\rm b}$, an increasing baryon-to-photon ratio decreases the pressure-to-density ratio of the plasma, altering the zero point of the acoustic oscillations. Modes that compress (into potential wells at the time of decoupling) compress more deeply, while those that rarify do not rarify as much. The net result is a boost to compression (odd) peaks and a suppression of rarefaction (even) peaks in the temperature power spectrum. Variation of $\omega_{\rm b}$ also affects the density of free electrons through recombination, and hence the damping scale. Both of these effects allow for tight constraints on the baryon density. 

\subsection{Applying the Ruler}
\label{sec:ApplyingRuler}
The amplitudes of the Fourier modes of density perturbations in the primordial plasma undergo damped and driven harmonic oscillation. Starting from rest, the solution in the radiation-dominated era well after horizon crossing is $\delta(k,\eta) \propto \cos(kr_{\rm s}(\eta) + \delta\phi(k))$. Since they start from zero initial momentum, the phase shift, $\delta \phi$, would be zero if it were not for the time-dependent driving caused by potential decay. Approximating projection from three dimensions to two as a mapping from $k$ to $\ell = k D_{\!A}^\star$, we find that the modes that give rise to the $p^{\rm th}$ peak, $k_p$, project to 
\be
\ell_p = k_p D_{\!A}^\star = p\left(\pi - \delta \phi(k_p)\right) D_{\!A}^\star/r_{\rm s}^\star.
\ee
Approximating $\delta \phi(k_p) = \delta \phi(k_{p+1})$ and defining $\Delta \ell = \ell_{p+1}-\ell_p$ we find
\be
\theta_{\rm s}^\star = \pi/\Delta \ell.
\ee
We thus see that the angular size of the sound horizon can be directly read off of the peak spacing \cite{hu1997a}. Although this is a good approximation, we note that the actual peak spacing differs somewhat due to details including geometric projection, gravitational lensing, contributions from velocity perturbations, breakdown of the tight coupling approximation, and the breadth of the visibility function. See \cite{pan2016} for a complete accounting.

With $r_{\rm s}^\star$ calculated, and $\theta_{\rm s}^\star$ inferred from the peak spacing, we can determine $D_{\!A}^\star\,{=}\,r_{\rm s}^\star/\theta_{\rm s}^\star$. 
In \lcdm, the comoving angular diameter distance to $z\,{=}\,z_*$ is related to energy densities via
\begin{align}
\label{eq:DA}
&D_{\!A}^\star = \int_0^{z_*}\frac{dz}{H(z)} =  2,998\;{\rm Mpc}\; \times \nonumber\\
&\int_0^{z_*} \dfrac{dz}{\sqrt{\omega_\Lambda + \omega_m (1+z)^3 + \omega_\gamma (1+z)^4 + \omega_\nu(z) }}.
\end{align}
We assume neutrino masses are specified \footnote{Note that increasing neutrino mass means increasing $\omega_{\nu}(z)$ and thus $\omega_\Lambda$ has to be reduced even further, driving $H_0$ down.} so the only remaining density we can adjust is $\omega_\Lambda$, which can be adjusted so that the model $D_{\!A}^\star$ is equal to the one inferred from $r_{\rm s}^\star$ and $\theta_{\rm s}^\star$. With this adjustment made, $H(z)$ is completely specified for all redshifts, including $z\,{=}\,0$, and therefore we have determined $H_0$.

\section{The $r_{\rm s}^{\rm drag}-H_0$ plane}

\begin{figure}
    \centering
    \includegraphics[width=\columnwidth]{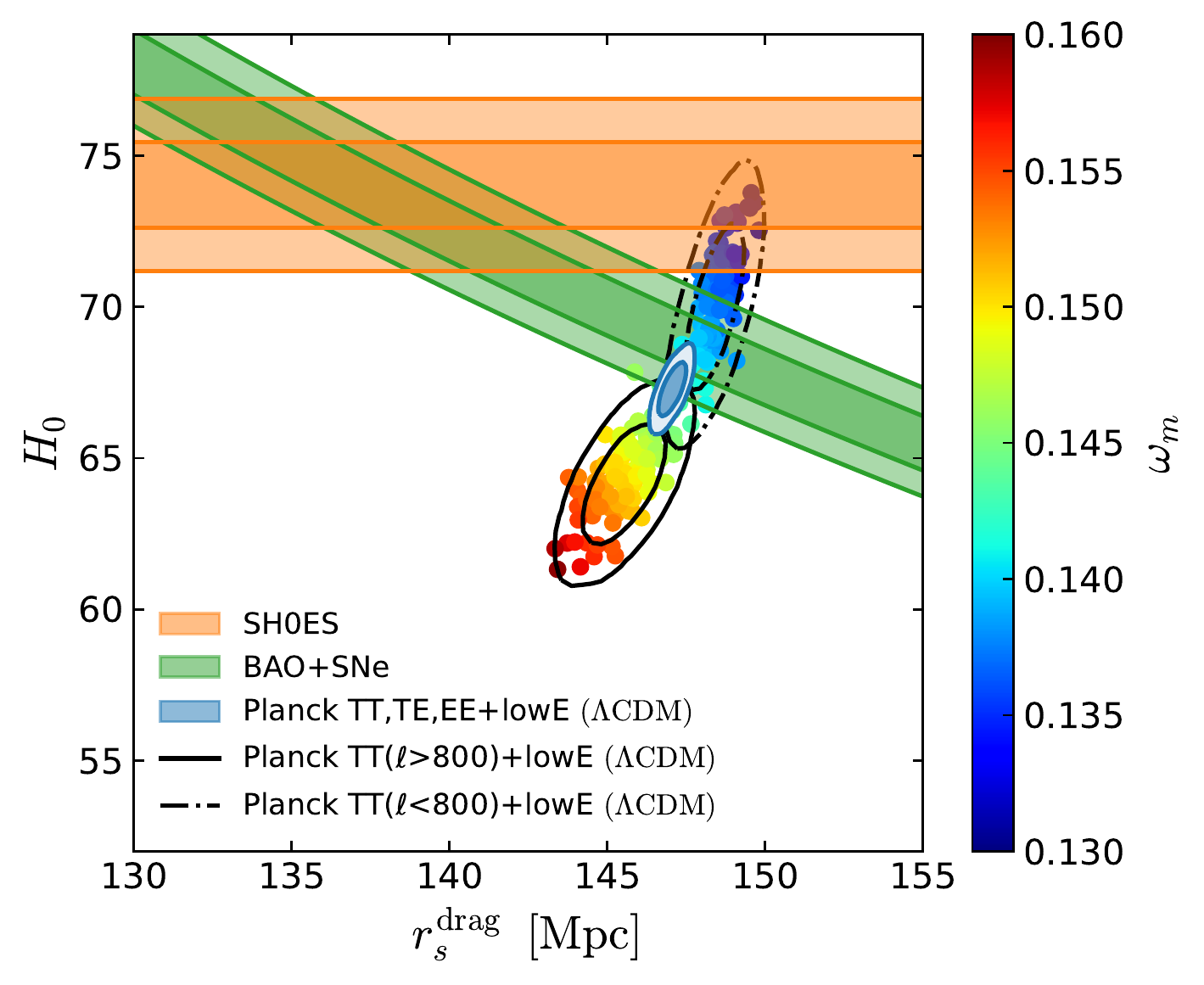}
    \caption{\LCDM tensions in the $r_{\rm s}^{\rm drag}-H_0$ plane. The orange and green shaded regions are 68\% and 95\% confidence regions from SH$_0$ES and from BOSS galaxy BAO + Pantheon, respectively. These inferences are largely independent of assumed cosmological model, as explained in the text. Conversely, the \planck contours assume \LCDM is the correct model at all redshifts. We show three versions of the Planck constraints: those from the full \planck TT,TE,EE+lowE likelihood, and those from \planck TT+lowE with TT limited to either $\ell\,{<}\,800$ or $\ell\,{>}\,800$. The color coding indicates values of the matter density, $\omega_{\rm m}$. We see a strong correlation between $\omega_{\rm m}$,  $H_0$, and $r_{\rm s}^{\rm drag}$. The direction swept out in the $r_{\rm s}^{\rm drag}-H_0$ plane by variations in $\omega_{\rm m}$ is not a direction that can reconcile all three datasets. 
}
    \label{fig:rs-H0}
\end{figure}

To understand the difficulty of reconciling CMB, BAO, and Cepheid-calibrated supernovae within \lcdm it is helpful to examine constraints in the $r_{\rm s}^{\rm drag}-H_0$ plane. In Fig.~\ref{fig:rs-H0} we show the Planck TT constraints from $\ell\,{<}\,800$ and from $\ell\,{>}\,800$, together with color coding of the mean value of $\omega_{\rm m}$. 

We see that increasing $\omega_{\rm m}$ leads to decreased $r_{\rm s}^{\rm drag}$ (and similarly decreased $r_{\rm s}^\star$), and also decreased $H_0$. This effect is straightforward to understand. A fractional change to $\omega_{\rm m}$ gives a fractional change in the sound horizon of $\fracdelta{r_{\rm s}^\star} \approx -\sfrac{1}{4}\,\fracdelta{\omega_{\rm m}}$ (this would be $\fracdelta{r_{\rm s}^\star} \approx -\sfrac{1}{2}\,\fracdelta{\omega_{\rm m}}$ in the absence of radiation, but the radiation softens the response \cite{hu2001}). To then keep $\theta_{\rm s}^\star$ fixed, we need to adjust $D_{\!A}^\star$ downward by the same fraction. The increase to the matter density does indeed serve to decrease $D_{\!A}^\star$, but by too much. To keep the distance from overshooting, $\omega_\Lambda$ must be adjusted downward. The net result is an $H(z)$ that is increased in the matter-dominated era, and decreased in the dark energy-dominated era, including a lower $H_0$ today. 

In Fig.~\ref{fig:rs-H0} we also show constraints from the SH$_0$ES distance ladder determination of $H_0$ \citepalias{riess2019} and from BOSS BAO plus Pantheon SNe distance measurements, both made without assumption of \lcdm. In place of the \lcdm assumption for the BOSS BAO plus Pantheon result we parameterize $H(z)$ with a spline with parameters controlling $H(z)$ at five points in redshift exactly as in \cite{bernal2016} and \cite{aylor2019}. To calculate a model $D(z)$ given a model $H(z)$ we assume zero mean curvature. The BAO points are measurements of $H(z)r_{\rm s}^{\rm drag}$ and $D(z)/r_{\rm s}^{\rm drag}$. The Pantheon data are uncalibrated, but constrain the shape of $D(z)$ (and thereby the shape of $H(z)$). The net result is we are able to extract a constraint on $\beta_{\rm BAO} \equiv c/(r_{\rm s}^{\rm drag} H_0)$ of $29.54\,{\pm}\,0.406$.

Although one can reduce the sound horizon within \lcdm by increasing the matter density, there is no value of the matter density that can simultaneously satisfy, within \lcdm, the distance ladder $H_0$, the BOSS + Pantheon constraint on $H_0 r_{\rm s}^{\rm drag}$, and Planck CMB data. Indeed, variation of the matter density takes the \lcdm prediction in a direction that is nearly orthogonal to the BAO + SNe constraint. These datasets thus severely restrict such variations, localizing them near the Planck prediction of $H_0$. Incidentally, this localization is why use of the inverse distance ladder tightens up the spread of \lcdm-predicted $H_0$ values coming from various CMB datasets. 

Note that the calibration of supernovae that leads to the \citetalias{riess2019} value of $H_0$ also leads to an empirical determination of $r_{\rm s}^{\rm drag}$ near 137 Mpc, as this is where their confidence regions overlap \cite{bernal2016,verde2017,aylor2019}. These datasets push us toward cosmological solutions that can lower the sound horizon inferred from CMB data.

\section{Post-recombination solutions}

This argument in favor of cosmological solutions that lower the sound horizon was made in \citetalias{aylor2019}. Implicit in this argument are some minimal cosmological assumptions necessary for the empirical determination of $r_{\rm s}^{\rm drag}$. In this section we consider a few potential solutions that violate these minimal assumptions and thereby do not require a reduced model sound horizon at recombination. 

We also revisit and improve an argument in \citetalias{aylor2019} against the possibility of post-recombination solutions. \citetalias{aylor2019} pointed out that the generation of new CMB anisotropies at late times, due to some beyond-\lcdm effects, could conceivably confuse our inference of cosmological parameters from CMB data. In such a scenario the sound horizon, inferred assuming \lcdm, could be larger than in reality. \citetalias{aylor2019} then argued against this possibility; we improve upon that argument here.

\subsection{High Sound Horizon Solutions}

\subsubsection{$H(z)$ Wiggles}

An example of a solution that reduces the tension with changes at low redshift is given by \citet{joudaki2018} and  \citet{keeley2019}. They both consider distance-redshift measurements at $z < 3$, including a BAO constraint from autocorrelation of flux transmission through the Lyman-$\alpha$ forest and cross correlation with quasars at $z \simeq 2.4$ \cite{dumasdesbourboux2017}, as well as constraints from Planck on the distance to the last-scattering surface. The constraints at $z=2.4$ on $D(z)/r_{\rm s}^{\rm drag}$ and $H(z)r_{\rm s}^{\rm drag}$, assuming \lcdm values for $r_{\rm s}^{\rm drag}$ are inconsistent with \lcdm at the 2 to 3 $\sigma$ level. \citet{keeley2019} consider two different models for the dark energy and find they can restore consistency. The tension with the BOSS BAO points remains though as long as they set $r_{\rm s}^{\rm drag}$ to the \lcdm value preferred by {\it Planck}, just as one would expect from the analyses in \citep{bernal2016,verde2017,aylor2019}.

Recently \citet{raveri2019} has explored a low-redshift reconstruction with even more degrees of freedom. He introduces, in an extension of the standard cosmological model, a large number of degrees of freedom affecting the expansion rate at redshifts between 0 and 9. One can think of this as a logical extension of the 5-point spline model used by \cite{bernal2016} and subsequently by \cite{aylor2019}, or as an extension of the approach in \cite{keeley2019}. Raveri's effective field theory approach ensures that the degrees of freedom in $H(z)$ include additional space and time-dependent perturbations that must also be present in a relativistically covariant theory. He finds, for the Scalar Horndeski model, an improvement in $\chi^2$ over the \lcdm case of 13.0, with 8.3 of that coming from the R18 likelihood. This improvement comes from a model space with 19.8 more effective degrees of freedom than in the \lcdm case. 

We see Raveri's work as opening up the possibility that there may be a late-time solution to the $H_0$ discrepancy; i.e., one that does not require a departure from \lcdm prior to recombination. It is interesting as an existence proof, but there are some things to keep in mind about the particular solution. First, it comes at the cost of a large number of new degrees of freedom; Raveri finds that most of the statistical evidence for any particular model is erased when including a penalty for the heavily widened prior parameter space. This is largely in line with \citet{poulin2018a}, who showed that a penalization technique known as "cross-validation" also disfavors several models which are qualitatively similar to Raveri's. Second, a critical aspect of the solution is some fast wiggles in $H(z)$, and therefore also in $D(z)$, in the redshift region of the three BOSS BAO redshifts. These wiggles might invalidate assumptions made in the reduction of BOSS data from a near-continuum of redshifts, to the publicly available constraints at three discrete redshifts.

\subsubsection{Violation of the distance duality relation}

The distance duality relation is the relation between luminosity distance and angular diameter distance. In any metric theory of gravity, and as long as photon number is conserved, the comoving angular diameter distance is related to the luminosity distance via $D_A(z) = (1+z) D_L(z)$ \cite{etherington33}. This relationship is assumed in our earlier discussion of empirical calibration of the sound horizon by the combination of Cepheid-calibrated supernovae and BAO angles. The mismatch between the empirical sound horizon and the \lcdm-determined sound horizon could arise due to a violation of one of the two assumptions underlying the distance-duality relation. See \cite{wu2015,ma2018a} for constraints on violations of the distance-duality relation from the combination of supernova and BAO data. 

For an example of a model in which such a violation occurs we point to the axion dimming phenomenon proposed by \citet{csaki2002} in 2002 as an acceleration-free means of explaining the dimness of high-redshift supernovae. In our case, Cepheid-calibrated supernovae appear too bright at redshifts $z \sim 0.6$ where there are the high-precision BOSS galactic BAO measurements. Rather than axion dimming, we appear to want axion brightening. This could be achieved if particles produced in a supernova explosion later converted to photons. If the amount of conversion is much less for more nearby supernovae, due to their shorter path lengths, then the Cepheid and BAO calibrations of supernova peak absolute magnitudes could be reconciled. 

This scenario faces a number of challenges. First, it does not a priori solve the problem of the CMB prediction of a lower $H_0$, nor does it solve the 3.1$\sigma$ tension between \lcdm\ and gravitational lensing time delays \cite{wong2019}. For those, we would also need $H(z)$ to depart from a \lcdm-compatible shape at $z < 0.6$. If we do that by varying the dark energy equation of state parameter, it will push it to less than -1. This has difficulties from a theoretical standpoint, and also makes the scenario seem complicated for explaining a supernova distance-redshift relation shape that is compatible with \lcdm in the absence of new physics.

Finally, there is also the challenge of building a viable particle physics model that would deliver an appropriate spectrum of photons. Can the photon coupling be sufficiently weak so as to prevent lack of neutrino production from SN 1987A, and yet strong enough for creation of sufficient photons over cosmological distances? Presumably the weakness of the interaction means the particles will decouple much deeper in the explosion than the photosphere and will therefore have a much hotter distribution than the photons coming directly from the photosphere.  \citet{meyer2017} claim that this is the case for axion production in core collapse supernovae, finding that for a 10 solar mass star, the resulting photon spectrum peaks near $\sim 100$ MeV. This is well above the energy of the photons relevant for the observations of Type Ia supernovae involved in SH$_0$ES measurement of $H_0$. We caution though that the above is for core collapse supernovae; we are unaware of a calculation of the spectrum from Type Ia supernovae. 
\comment{The spectrum may be expected to peak at smaller energies for Type 1A, although we are not aware of detailed predictions in the literature. (TBD for Lloyd to remove)
}

\comment{
Conversion of photons to other particles, or  A second possible means of eliminating tension via a purely late-time cosmological solution could come from including new photon interactions that would cause non-geometric altering of the supernova luminosity-distance-flux relationship. Axion dimming is one way to alter this relation -- an effect that the authors of \cite{csaki2002} proposed, in 2002, as a cosmic-acceleration-free means of explaining anomalously dim supernovae.  

Axion dimming reduces the flux of supernovae, and preferentially so for more distant supernovae, as, in general, the further the light has traveled, the greater the optical depth to axion conversion, asymptoting to 33\% due to equilibration with axion-to-photon conversion. One can place constraints on the photon-axion coupling by comparing luminosity distances and angular-diameter distances to the same redshift (cite needed).

This effect, however, goes in the opposite direction of what is needed to reduce the $H_0$ tension. If we corrected for axion dimming, we would find the supernovae to be closer and thus $H_0$ to be higher. Further, the amplitude of this effect is severely constrained by the shape of the distance-redshift relation out to $z \sim 1$; if there is such an effect it must have a negligible impact on $H_0$ inference.

An approach, like the axion dimming one, that modifies the purely geometric dimming of supernovae, has to either preferentially dim the supernovae in the Cepheid calibration sample, or preferentially brighten the supernovae in the Hubble flow. The former option has the challenge that the supernovae would have to be differentially impacted relative to the Cepheids, or the effect would cancel out in the calibration process. The latter option, requiring a brightening of supernovae, seems to require that other particles emitted by the supernovae convert to photons. We find this prospect to be unlikely. Even if such a mechanism were found, in general one would expect it to have significant impact on the shape of the supernova distance-redshift relation. The consistency of that shape with \lcdm would thus appear to be a conspiracy. Non-geometric flux effects appear to be a challenging way to solve the Hubble tension.

}

\subsubsection{Cepheid (mis)-calibration}

\citet{desmond2019} consider the possibility of a fifth force that impacts the Cephied period-luminosity relation in an environmentally-dependent manner. The fifth force is a long-range force that augments the gravitational force. It is "screened" at sufficiently high densities as to avoid solar system constraints on such fifth forces. They argue that some of the Cepheids used for calibrating supernovae may be unscreened while the LMC, Milky Way and NGC 4258 Cepheids that are calibrated via geometrical methods, are screened. The unscreened Cepheids effectively experience a larger gravitational constant $G$, that alters the period-luminosity relationship. The net result is a bias in supernova calibration resulting in an increase in the inferred $H_0$. 

The impact of some new physics on Cepheid dynamics is a logical possibility for a reduction in $H_0$ discord. The \citet{desmond2019} paper is the only proposal of this kind of which we are aware. In general, and in particular with \cite{desmond2019}, such solutions do not impact the 3.1$\sigma$ tension between \lcdm\ and gravitational lensing time delays \cite{wong2019}.

\subsection{Low Sound Horizon Solutions}

We consider here two distinct paths toward lowering the sound horizon with only post-recombination changes to the cosmological model. The second of these is not aimed at lowering the sound horizon at recombination, but rather at reducing the size of the comoving feature in the galaxy 2-point correlation function, and thereby lowering the model prediction for the distance-ladder-determined sound horizon. We see neither of these as likely paths toward a viable solution. Our discussion here serves to support the case that lowering the sound horizon requires changes to the cosmological model that are important {\em prior} to recombination.

\subsubsection{Late-time Confusion Sowing}
\label{sec:late-time-confusion}

To reduce the sound horizon without a departure from the \lcdm model prior to recombination, the matter and baryon densities, which uniquely control the sound horizon in \lcdm, must actually be different than what we have inferred by fitting the \lcdm model to CMB data. This could arise because departures from \lcdm after recombination have confused our interpretation of the CMB data. 

We can immediately eliminate the possibility of a confused $\omega_b$ resolving the tension. We would have to nearly double the baryon density to achieve a 7\% reduction in $r_{\rm s}^\star$. This would be severely at odds with the baryon density inferred, using minimal late-time cosmological assumptions, from primordial elemental abundances \cite{aver2015,cooke2018}.

A confusion in the determination of $\omega_{\rm m}$ seems initially, at least, more plausible. We can think of three post-recombination effects that can alter or create new CMB anisotropies and cause this confusion: deflection due to spatially varying gravitational potentials (gravitational lensing), anisotropies induced by time-varying potentials (the Integrated Sachs Wolfe (ISW) effect), and geometric projection, which could be altered by changing the angular-diameter distance to recombination. 

We view the first two of these to be exceedingly unlikely solutions. Regarding ISW, \citetalias{aylor2019} report that {\it Planck} EE and TE spectra give an $r_{\rm s}^{\rm drag}$ that is larger than the distance ladder determination by  $2.3\,\sigma$ and $2.7\,\sigma$ respectively. There is no ISW effect on polarization, so these results are unaffected by model changes that alter anisotropies solely through ISW effects. 

For gravitational lensing, the impact on the CMB spectra is determined by the lensing potential power spectrum, upon which there are already strong constraints based on reconstruction from the CMB four-point functions \cite{adhikari2019}. The reconstructions of the lensing potential power spectrum are fairly model independent, hence any effort to introduce significant changes to the CMB spectra from non-standard lensing have to take these into account. Changes to the CMB spectra, resulting from lensing changes, respond more slowly than the lensing power spectrum itself, and lensing power measurements have sufficient sensitivity to determine the amplitude of the lensing power to better than 3\% \cite{planck2016-l08}. 

Our third post-recombination effect is simply a change to the angular-diameter distance. But there is no change to the angular-diameter distance that can bring down $r_{\rm s}^\star$ (by increasing $\omega_{\rm m}$) and keep both $\theta_{\rm s}^\star$ and $\theta_{\rm s}^{\rm EQ}$ within acceptable ranges for the data.

The reason the CMB data cannot tolerate much of a change to $\theta_{\rm s}^{\rm EQ}$ has to do with the resonant enhancement of acoustic oscillation amplitudes that occurs for modes that begin oscillating in the radiation-dominated era, due to gravitational potential decay, the radiation-driving phenomenon of \citet{hu1997} that we discussed in Section~\ref{sec:calibrate_ruler}. Modes that begin oscillating closer to matter-radiation equality suffer less potential decay and therefore enjoy less of a boost in their amplitudes. In the \lcdm model the radiation-driving envelope is parameterized by one number: $\theta_{\rm s}^{\rm EQ}$. Defining $\ell_{\rm s}^{\rm EQ} \equiv k_{\rm s}^{\rm EQ} D_{\!A}^\star = 1/\theta_{\rm s}^{\rm EQ}$~\footnote{The absence of a factor of $\pi$ in the relationship between $\ell$ and the corresponding $\theta$ here has to do with the definition of $\theta_{\rm s}^{\rm EQ}$ as a projection of $1/k_{\rm s}^{\rm EQ}$ rather than a half wavelength}, the Planck data tell us $\ell_{\rm s}^{\rm EQ} \simeq 223$.

Assuming \lcdm prior to matter-radiation equality we have 
\be
k_{\rm s}^{\rm EQ} \equiv 1/r_{\rm s}^{\rm EQ} \simeq \sqrt{3} \rho_{\rm m} \sqrt{\frac{16\pi G}{3\rho_{\rm rad}}}
\ee
where $\rho_{\rm m}$ and $\rho_{\rm rad}$ are the matter and radiation densities that we would have today if these were to scale with redshift as $(1+z)^3$ and $(1+z)^{4}$ respectively, from the epoch of equality, and $k_{\rm s}^{\rm EQ}$ is the comoving wave number of a mode that "crosses the (sound) horizon" at matter-radiation equality. We see from this that $k_{\rm s}^{\rm EQ} \propto \omega_{\rm m}$. 

Suppose then that we reduced $r_{\rm s}^\star$ by 7\% by increasing the matter density according to $\fracdelta{r_{\rm s}^\star} \simeq -\sfrac{1}{4}\,\fracdelta{\omega_{\rm m}}$, while simultaneously reducing $D_{\! A}^\star$ by the same amount via some new physics so as to keep $\theta_{\rm s}^\star$ fixed. However, in doing so we have also changed $\fracdelta{\ell_{\rm s}^{\rm EQ}}$ by $-\sfrac{1}{4}\,\fracdelta{\omega_m} + \fracdelta{\omega_{\rm m}} = 3\,\fracdelta{r_{\rm s}^\star}$. For the 7\% change in $r_{\rm s}^\star$ necessary, this is a shift in $\ell_{\rm s}^{\rm EQ}$ that is not reconcilable with the CMB power spectra determinations. The shift in the matter density would also impact the scale of the turnover in the matter power spectrum, a scale that itself responds four times more rapidly than does $r_{\rm s}$ \cite{sutherland2012a}.

We conclude that if we are to have a cosmological model solution to the $H_0$ problem, it is highly likely that it would include departures from the standard cosmological model prior to recombination. To reduce the sound horizon, these changes have to be important near recombination, as \citetalias{aylor2019} pointed out, and as we will see shortly from Fig.~\ref{fig:integrand}.

\subsubsection{Post-recombination evolution of $r_{\rm s}^{\rm drag}$}

Evolution of the BAO feature in the galaxy two-point correlation function toward smaller comoving distance between recombination and its observation at $z \lesssim 1$ could, in principle, reconcile the model $r_{\rm s}^{\rm drag}$ with its empirical determination. 

Such a scenario almost certainly requires new physics. The relevant length scales are sufficiently large that we can expect perturbation theory to be an accurate approximation. Fluctuations are less than unity in the mass density today when smoothed on scales larger than about 10 Mpc. Non-linear evolution, redshift-space distortions, and galaxy biasing effects are important at the sub-percent level, e.g., \citep{padmanabhan2009}. These effects are small, and taken into account in the recovery of the angular and redshift acoustic scales from the galaxy two-point correlation function. Use of an especially robust feature in the correlation function, the so-called linear point, achieves similar results \citep{anselmi2018}.

Perhaps a modification of gravity, making it stronger on large scales, could alter the Green's function, increasing the rate of infall of the shell of matter around an initial overdensity toward the location of the initial overdensity. However, for the peak to move inward would require transport of mass, which, if the process were still on-going today, and had not somehow come to a halt by $z < 1$, would result in much larger peculiar velocities than predicted by standard theory. 

The peculiar velocities associated with this infall have not been observed. The observed redshift-space maps are consistent on large scales with the expected relationship between the density and velocity fields. 
The success of BAO peak reconstruction \citep{eisenstein2007} argues that standard theory accounts for the relationsip between the density field and the peculiar velocities. In this process, large-scale peculiar velocities are estimated from the observed redshift-space distribution of galaxies, using perturbation theory. These velocities are then used to reconstruct the original density field. This process is intended to sharpen up the acoustic feature in the correlation function, which gets smeared out by transport. Its success in application to data e.g., \citep{alam2017} argues that we understand transport of mass on the relevant length scales, and argues against any models in which this is significantly different.

\section{Pre-recombination solutions}

In this section we consider cosmological solutions that depart from \lcdm prior to recombination. In general they have departures after recombination as well. 
We group them into four categories: confusion sowing (Subsection~\ref{sec:confusionsowing}), sound speed reduction (Subsection~\ref{sec:soundspeedreduction}), high-temperature recombination (Subsection~\ref{sec:hightemprec}), and increased $H(z)$ (Subsections ~\ref{sec:photoncooling} and~\ref{sec:addcomponents}). 

Models in the confusion sowing category sow confusion in parameter determination in the same sense as in the late-time confusion sowing scenario. The difference is that here we have concrete examples as there are many in the literature, and they all include model changes away from \lcdm that are important prior to recombination.

\subsection{Confusion Sowing}
\label{sec:confusionsowing}

Models in this category have the same matter content as \lcdm: radiation, non-relativistic matter, and a cosmological constant. The differences are in initial conditions or some additional interactions (ones which do not change the sound speed of the plasma or substantially alter recombination, as these are discussed separately in other subsections). These differences change the value of $\omega_{\rm m}$ inferred from data, from what it would be if the inference were done assuming \lcdm; i.e., if the model is correct, then the beyond-\lcdm aspects of the model have {\em confused} the \lcdm-based inference of $\omega_{\rm m}$. Examples include the interacting neutrino model \citep{cyr-racine2014a,lancaster2017}, the modified gravity model of \cite{lin2019a}, the introduction of extra freedom in the primordial power spectrum, and the super-sample covariance model of \cite{adhikari2019}.

We find this category of interest because we can show that models in this category cannot produce a complete reconciliation of CMB, Cepheid, supernovae, and galactic BAO data. Assuming the models do not confuse our interpretation of $\theta_{\rm s}^\star$ substantially, then the predictions of $r_{\rm s}^\star$ and $H_0$ are both controlled by one parameter: $H_0$. We have already seen in Fig.~\ref{fig:rs-H0} that fluctuations in $\omega_{\rm m}$ move the predictions in the $r_{\rm s}^{\rm drag}-H_0$ space in a direction that either improves agreement with BAO + uncalibrated SNe, or with Cepheids + SNe, but not both.

Placing the interacting neutrino case in this category comes with a caveat as these models do have an important impact on our inference of $\theta_{\rm s}^\star$ -- large enough to have a significant influence on the inference of $H_0$ even at fixed $\omega_{\rm m}$. The interactions reduce or eliminate the free-streaming-induced temporal phase shifts of the acoustic oscillations \cite{bashinsky2004}. These shifts alter peak locations, have been detected in the CMB temperature power spectrum \cite{follin2015,baumann2016}, and affect our inference of $\theta_{\rm s}^\star$ as described in Section~\ref{sec:ApplyingRuler}. This impact on $\theta_{\rm s}^\star$ inference from the free-streaming-induced phase shift was first noted in \cite{hou2014} with the combination of WMAP7 \cite{larson2011} and SPT-SZ survey data \cite{story2013}. With Planck temperature and polarization data \citet{kreisch2019} found that the reduction in phase shift caused by neutrino interactions led to a $> 10\,\sigma$ upward shift in $\theta_{\rm s}^\star$ of 0.5\%. Variation of the cosmological constant to achieve a corresponding 0.5\% decrease to the distance to last scattering, at fixed $\omega_{\rm m}$, increases $H_0$ by 2.6\%, a non-negligible shift. 

\subsection{Sound Speed Reduction}
\label{sec:soundspeedreduction}
The adiabatic sound speed, $c_s$, is related to the baryon-photon plasma density and pressure via $c_s^2 = \partial P/\partial \rho$. For the baryon-photon plasma, the inertia of the baryons reduces the pressure from the pure relativistic gas case of $P=\rho/3$ to $P=\rho/(3(1+R))$ where $R=3\rho_{\rm b}/(4\rho_\gamma)$. The sound speed could potentially be reduced further by introduction of a new non-relativistic species tightly coupled either to the photons or to the baryons, species $x$, so that $R$ becomes 3$(\rho_{\rm b}+ \rho_x)/(4\rho_\gamma)$. However, this species $x$ would, due to this reduction of pressure (at fixed photon density), contribute in the same way that baryons do to the odd-even peak height modulation. Thus if species $x$ is really there, we already have its influence included in to our estimate of the sound speed from CMB data. 

Constraints on dark matter--proton interactions have been studied in \cite{boddy2018,boddy2018a}.  \citet{boddy2018} studied such a scenario and found that in the case with a fraction of the dark matter strongly coupled to baryons and dark matter, the fraction had to be less than 0.4\% to be consistent with the 2015 Planck temperature, polarization, and lensing data.  This upper limit is sufficiently low that these interactions have negligible impact on the sound speed compared to the 7\% discrepancy in \lcdm and empirically-determined sound horizons. Further, we expect anti-correlation between this fraction and the baryon-to-photon ratio, for reasons given in the preceding paragraph, that further reduce the impact on the sound speed.

\subsection{High-temperature Recombination}
\label{sec:hightemprec}
The remainder of our solutions are all ways to reduce the conformal time to the end of the baryon-drag epoch, 
\be
\eta_{\rm d} = \int_0^{t_{\rm d}} \frac{dt}{a(t)} = \int_{z_{\rm d}}^\infty \frac{dz}{H(z)},
\ee 
which, ignoring time-dependence of the sound speed gives the comoving sound horizon via $r_{\rm s}^\star = c_{\rm s} \eta_{\rm d}$. In this subsection we consider reducing this conformal time by reducing $z_{\rm d}$ by having the baryon drag epoch end at a higher photon temperature. Such a solution was in fact presented by \cite{chiang2018}. However, the question remains of the underlying physics that would lead to a high-temperature recombination.

In principle, it could be achieved with time variation of the fine structure constant, since a stronger electromagnetic interaction would lead to recombination at a higher temperature. Based on CMB power spectra \cite{hart2018} find the change in the value of $\alpha$ between recombination and today to be $\fracdelta\alpha = (.7 \pm 2.5)\times 10^{-3}$. Since atomic physics energies are linearly proportional to $\alpha$, this indicates only sub-percent changes in recombination temperature are permissible. These are too small to achieve a 7\% change in the sound horizon.

The failure of $\alpha$ variation as a way to get to small $r_{\rm s}^\star$ is a specific example of what we expect to be true in general: changes to the physics of recombination sufficient to change the sound horizon by 7\% will wreak havoc on the shape of the damping tail. Admittedly, we have no proof that such a solution is not possible. But it seems highly unlikely that new physics alters $r_{\rm s}^\star$ by changing recombination, while having an acceptably small impact on the shape of the CMB damping tail.

The unlikeliness is underscored by the fact that recombination occurs out of chemical equilibrium -- the relevant atomic per-particle reaction rates are not much faster than the Hubble rate. The particular details of the ionization history resulting from this out-of-equilibrium recombination are marvelously consistent with the shape of the damping tail. Thus the task is more challenging than simply reproducing a generic equilibrium ionization history at a higher temperature.

\subsection{Photon Cooling / Conversion}
\label{sec:photoncooling}

The conformal time to recombination could also be reduced if the photons cooled more rapidly than adiabatically just prior to recombination. Some unknown species, via some unknown (and previously ineffective) interaction, could cool the photons, so that their temperature drops more rapidly than in the absence of such cooling. 

\cite{chiang2018} point out that spectral distortions, and measurements of the CMB spectrum, constrain this manner of solution. Since photon number decreasing and increasing reactions are slow at $z \lesssim 10^7$, we generically expect CMB photon cooling to lead to observable spectral distortions. They also point out that this can be avoided if all the action is sufficiently far out on the Wien tail. This is the region of the spectrum that needs to be affected to push recombination back earlier, and it is also a region of the spectrum that is much less well constrained by observations of the CMB spectrum.  

However, the shape of the damping tail {\em is} impacted exactly by these Wien tail photons. Again, we find it exceedingly unlikely that the observed damping tail shape consistency (with the standard model non-equilibrium calculation) follows from a coincidence.

\subsection{Increasing $H(z)$ with additional components}
\label{sec:addcomponents}
We consider several different types of additional contributors to the energy density that could increase $H(z)$ just prior to recombination.

The increased expansion rate leads to two physical effects, both of which decrease $r_{\rm s}^\star$. First, there is the reduction in conformal time required to cool to a given temperature. Second, we find that the temperature at last-scattering (i.e. $z_\star$) increases very slightly, as we discuss in Appendix~\ref{app:visrsrd}. The former is, by far, the dominant effect.

To understand the challenges faced by model solutions, like \lcdm{+}$N_{\rm eff}$, that increase $H(z)$ prior to recombination, it is helpful to think of their impact on three angular scales: $\theta_{\rm s}^{\rm EQ}$, $\theta_{\rm s}^\star$, and $\theta_{\rm d}^\star$.  These are, respectively, the angular sizes of these length scales projected to today from the last-scattering surface: the comoving size of the sound horizon at matter-radiation equality $r_{\rm s}^{\rm EQ}$, the comoving size of the sound horizon at recombination $r_{\rm s}^\star$, and the comoving size of the photon diffusion scale at recombination $r_{\rm d}^\star$. The importance of these scales for thinking about light relics has been emphasized previously in \cite{hu1997,bashinsky2004,hou2013}.

To keep $\theta_{\rm s}^\star$ from changing too much we have $\fracdelta{D_{\!A}^\star} \simeq \fracdelta{r_{\rm s}^\star}$. To keep all three angular scales thus requires the ratios of their associated length scales to also not change by too much.

The data also do not tolerate too much of a change to the photon diffusion scale. Photon diffusion has a huge impact
on the amplitudes of the power spectra, reducing fluctuation power by a factor of $\sim 30$ at $\ell = 2000$.  The impact of photon diffusion is largely, though not entirely, captured by the photon diffusion scale, computed to second order in tight coupling, $r_{\rm d}^\star = \pi/k_{\rm d}$ \cite{kaiser1983}. From {\it Planck} data we have $\ell_{\rm d} \equiv k_{\rm d} D_{\!A}^\star \simeq 1950$. How much a model can depart from this $\theta_{\rm d}^\star$ depends on how well other parameter variations can mimic the impact of photon diffusion. 

It is conventional to calculate $r_{\rm s}^\star$ and $r_{\rm d}^\star$ via integrals from very early times to the midpoint of recombination. These choices are all artificial to some degree. Here we introduce averaged quantities, $\bar r_{\rm s}$ and $\bar r_{\rm d}$, that remove some of the arbitrariness of the choice of a particular redshift. See the Appendix for details.

In Fig.~\ref{fig:integrand} we show how $\bar r_{\rm s}$ and $\bar r_{\rm d}$ respond to changes in $H(z)$. The left vertical axis tells us the fractional change in length scale for a given fractional variation in $H(z)$ per logarithmic redshift interval. Notice that, as pointed out in \citetalias{aylor2019} the sound horizon is most sensitive to the expansion rate in the decade of expansion just prior to recombination. In contrast, the diffusion scale sensitivity to expansion rate is more compactly contained near recombination. We will refer to the model curves and the vertical axis further below.

Fig.~\ref{fig:integrand} does not include a similar response for $r_{\rm s}^{\rm EQ}$ to variation in $H(z).$ The reason is that as soon as we have components contributing to $H(z)$ that are neither matter nor radiation, the length scale becomes poorly defined. What matters physically is the radiation-driving envelope which, in general, is not parametrized by just one number, and whose shape will change away from the \lcdm shape as $H(z)$ is varied prior to recombination. As we discuss in Section~\ref{sec:omegamscatter}, the oscillatory residuals in the Planck data, fit to \lcdm, that drive the high $A_{\rm L}$ and angular-scale-dependent values of $\omega_{\rm m}$, might find an explanation in a model with a different shape to the radiation-driving envelope. 

With these preliminary remarks out of the way, we are now ready to discuss particular approaches to increasing $H(z)$ by adding new components, beginning with additional light relics.

\subsubsection{Additional thermal relativistic species}

Additional light degrees of freedom are ubiquitous in extensions of the standard model of particle physics. Thermally produced, relativistic species are an extremely-well motivated extension of \lcdm to consider, and one that will increase $H(z)$ in the necessary window in redshift. It is thus not surprising that increasing $N_{\rm eff}$ (which is the common parameter for quantifying an increase in energy density from relativistic particles) is a well-considered way to improve agreement between $H_0$ from CMB data with $H_0$ from Cepheids + SNe; e.g., \cite{eisenstein2004}.

\citet{hou2013} describe how other parameters shift in order to still fit the CMB data as we allow $N_{\rm eff}$ to vary. To keep variation in the redshift of matter-radiation equality within the tight range allowed by the CMB data,  the matter density must scale just like the radiation density, i.e. by $1+R_\nu$ where $R_\nu\,{\equiv}\,\sfrac{7}{8}(\sfrac{4}{11})^{\sfrac{4}{3}}N_{\rm eff}$ is the ratio of non-photon radiation energy density to photon energy density. This then implies that $r_{\rm s}^\star\propto(1+R_\nu)^{-\sfrac{1}{2}}$, meaning that $N_{\rm eff}$ would have to increase to around $N_{\rm eff}\,{=}\,4.2$ to achieve a 7\% reduction in $r_{\rm s}^\star$.

To keep $\theta_{\rm s}^\star$ fixed while changing $r_{\rm s}^\star$, we must also have $D_{\!A}^\star \propto r_{\rm s}^\star \propto (1+R_\nu)^{-\sfrac{1}{2}}$. Since the matter density is already varying in a way to provide this scaling, the other dominant component, $\rho_\Lambda$, must as well. Neglecting effects of neutrino mass, we thus also have the low-redshift $H(z)$ (and therefore $D_A(z)$) also changing amplitude but not shape. Thus an increased $N_{\rm eff}$ leads to decreased sound horizon, increased $H_0$, and almost no change to $H(z)r_{\rm s}^\star$ and $D(z)/r_{\rm s}^\star$ and hence to the BAO observables. 

\begin{figure}[!tbh]
\includegraphics[width=\columnwidth]{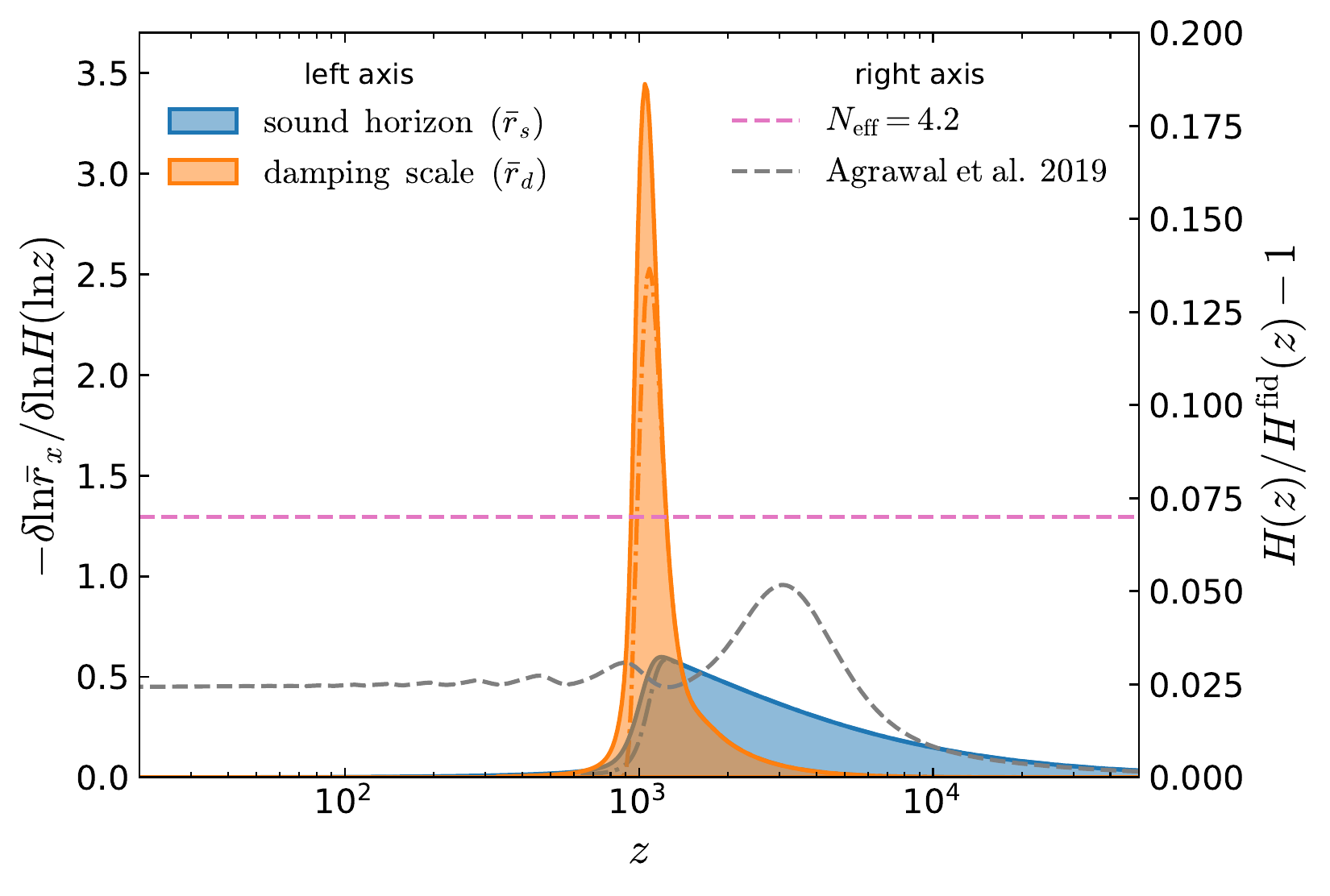}
\caption{On the left axis (the filled curves), we show the fractional linear response of the "visibility-averaged" $\bar r_{\rm s}$ and $\bar r_{\rm d}$ to a fractional change in $H(z)$ in some logarithmic interval in $z$ (see Appendix~\ref{app:visrsrd} for exact definitions). For each curve, the dot-dashed line shows what the response would be without accounting for the dependence of the visibility function on $H(z)$. The right axes (dashed curves) show the fractional change in $H(z)$ relative to our \lcdm fiducial model for two cases which reduce $\bar r_{\rm s}$. The first has $N_{\rm eff}\,{=}\,4.2$ (which lowers $\bar r_{\rm s}$ by 7\%) and the second is the best-fit $\phi^4$ model from \citet{agrawal2019}. One can read off the (linearized) change to $\bar r_{\rm s}$ and $\bar r_{\rm d}$ from these two models by multiplying the dashed lines by either the blue or orange regions, respectively, then integrating across $z$.}
\label{fig:integrand}
\end{figure}

Because the shape of $H(z)$ does not change, we then have the simple relation
$\fracdelta{r_{\rm d}^\star} \simeq \sfrac{1}{2}\,\fracdelta{r_{\rm s}^\star}$. One can also infer this relation from Fig.~\ref{fig:integrand}. With the dashed magenta line, we show the change to $H(z)$ from increasing $N_{\rm eff}$ to 4.2 while holding $r_{\rm s}^{\rm EQ}$ constant. Multiplying this line by either of the $\bar r_{\rm s}$ or $\bar r_{\rm d}$ visibility-averaged (dot-dashed) contours and integrating across redshift gives the total linearized change to these parameters due to the change in $H(z)$. As the $H(z)$ change is, in this case, constant, this is just the total area under each curve. Indeed, the area is about twice as large under the blue curve vs. the orange curve, as expected from the discussion above.

We thus find that a change to $N_{\rm eff}$ ends up affecting the ratio of $r_{\rm d}^\star/r_{\rm s}^\star$, which is undesirable as the observed $\theta_{\rm d}^\star/\theta_{\rm s}^\star$ is consistent with the standard model value. The unavoidable altering of $\theta_{\rm d}^\star/\theta_{\rm s}^\star$, in the case of \lcdm{+}$N_{\rm eff}$ is one of two main reasons the CMB data do not prefer larger values of $N_{\rm eff}$. The other is the shift in the temporal phase of the acoustic oscillations and the associated shifts in acoustic peak and trough locations \cite{follin2015,baumann2016}. 

Additional interactions between the light relics, or between the light relics and the dark matter, make the phenomenology more complicated and, as has been shown in, for example, \citet{kreisch2019}, can increase the value of $N_{\rm eff}$ that the CMB data can tolerate. With a dataset combination that includes R19, \citet{kreisch2019} find $N_{\rm eff} = 4.02 \pm 0.29$ (as well as the $H_0 = 72.3 \pm 1.4$ \hubbleunit\ mentioned in the introduction.) \citet{blinov2019} explore constraints from laboratory experiments on models that can deliver the neutrino-neutrino scattering cross sections desired by these fits to data, that are larger than those in the standard model by 8 orders of magnitude. While they find highly significant constraints, not every possibility is ruled out. A light relic solution to the Hubble tension remains an intriguing possibility. 

\subsubsection{Early Dark Energy}
Models that also change the shape of $H(z)$, such as early dark energy models \citep{doran2006,bielefeld2013,karwal2016}, have a chance of avoiding this fate, since the $r_{\rm s}^\star$ and $r_{\rm d}^\star$ integrands are different. The extra component in these models is a scalar field that, at least temporarily, near recombination, behaves like a cosmological constant. More recently \citet{poulin2018,agrawal2019} and \citet{lin2019} considered a class of scalar field models due to the expectation that they might be able to solve the $r_{\rm d}^\star/r_{\rm s}^\star$ problem faced by light relics. 

The model curves in Fig.~\ref{fig:integrand} are for a \lcdm{+}$N_{\rm eff}$ model and for one of the \citet{agrawal2019} best-fit $\phi^4$ model. The \lcdm{+}$N_{\rm eff}$ model has, in addition to the increase in $N_{\rm eff}$, an increase in $\omega_{\rm m}$ that preserves $r_{\rm s}^{\rm EQ}/r_{\rm s}^\star$ by design. As we discussed in the previous subsection, this model delivers $\fracdelta{r_{\rm d}^\star} \simeq \sfrac{1}{2}\,\fracdelta{r_{\rm s}^\star}$, and so $r_{\rm d}^\star$ does not decrease enough to preserve $r_{\rm d}^\star/r_{\rm s}^\star$. The \citet{agrawal2019} model does not remedy this challenge. In fact, since more of its change to $H(z)$ is under the blue rather than the orange curve, the $r_{\rm d}^\star/r_{\rm s}^\star$ ratio is even more affected, leading to additional (seemingly unwanted) damping in the power specrum at fixed $\ell$. 

Nevertheless, the \citet{agrawal2019} model provides a better fit to the CMB data. The increase in damping is partially compensated by an increase in $n_{\rm s}$ and shifts in other parameters. Thus it is a demonstration that the data can tolerate a model with a fairly large departure of $r_{\rm d}^\star/r_{\rm s}^\star$ from its \lcdm value, as also seen in \cite{poulin2018}. We suspect that the \citet{agrawal2019} solution is being driven more by the impact of the increased expansion rate on the amplitude of acoustic oscillations via the radiation driving effect. Similar conclusions were reached in \cite{poulin2018}. It may be explaining the residuals in the fit to \lcdm that lead to the anomalously large $A_{\rm L}$ \cite{planck2016-l06} and somewhat inconsistent values of the matter density inferred from different angular scales \cite{planck2014-a13,addison2015,planck2016-LI}. We will discuss these possibilities more in
Section~\ref{sec:omegamscatter} below. 

While we have chosen the specific example of $\phi^4$ as it was the monomial potential found by \cite{agrawal2019} to best alleviate tension among the combined datasets, we should point out that \citet{lin2019a} and \citet{smith2019} find that Planck data are better accommodated by a potential that flattens at high field value, a preference primarily driven the polarization.

\subsubsection{Designer $H(z)$}

\citet{hojjati2013} study parameterized departures from $H(z)$ in the standard model between $z=10^5$ and $z=0.1$ with an effective dark component with several different choices of the sound speed, as constrained by the first release of {\it Planck} data. The upper limits to the fractional perturbations to $H(z)$ are at the 4 to 10\% level depending on $a$ and sound speed. Such changes may be sufficient to reduce the sound horizon by the 7\% required for optimal agreement with the CDL. It would be interesting to redo their analysis with the {\it Planck} 2018 data, including polarization, and to determine the resulting constraints on $r_{\rm s}^\star$. 

\section{Excess Scatter in Inferences of Matter Density Across Angular Scale}
\label{sec:omegamscatter}

Above we discussed the angular scale $\theta_{\rm s}^{\rm EQ}$ and the desire to preserve it, since this scale controls the radiation-driving envelope. However, in general, the shape of the radiation-driving envelope depends on more than just this one number. The impact on acoustic oscillation amplitude from gravitational potential decay varies with perturbation wavelength, or angular scale, quite slowly, since mean matter and radiation densities redshift differently only by one power of the scale factor. There is information in this envelope about the history of the expansion rate over a wide range of redshifts. 

If $H(z)$ is varied by an overall rescaling, such as by an increase in the radiation density and the matter density in proportion, then $\theta_{\rm s}^{\rm EQ}$ does not change, and the entire shape of the radiation-driving envelope is also unchanged. In general though, a change to the shape of $H(z)$ in the decade of scale factor evolution prior to recombination will alter the shape of the radiation-driving envelope, even if $\theta_{\rm s}^{\rm EQ}$ is kept fixed. 

How would these radiation-driving envelope shape changes show up in analyses of CMB data? The primary source of information about the matter density at $\ell < 1400$ or so~ \footnote{At smaller scales the impact of $\omega_{\rm m}$ on lensing becomes an important source of information about $\omega_{\rm m}$ \citep{planck2016-LI}.} stems from its impact on the radiation-driving envelope. Therefore, if these analyses are done assuming \lcdm then radiation-driving envelope shape changes would likely show up as excess scatter in the matter density, $\omega_{\rm m}$, inferred across different angular scales.

There is some evidence of this excess scatter. \citet{addison2015} compared parameters inferred from Planck TT power spectra at $\ell < 1000$ with $\ell > 1000$ and found $\omega_{\rm c}$ discrepant at $2.5\,\sigma$. The Planck Collaboration found, in \cite{planck2016-LI}, for the same data set and the same angular-scale split, that $\omega_{\rm m}$ was the most discrepant of the five parameters evaluated, although with a slightly lower statistical significance at $2.3\,\sigma$~\footnote{The lower significance is due to a more accurate calculation that used simulations to avoid some approximations. We also note that \cite{planck2016-LI} showed that the significance of finding any one of the parameters this discrepant or more is only $1.6\,\sigma$.}. The dependence of the inferred $\omega_{\rm m}$ on chosen angular scale can also be seen in our Fig.~\ref{fig:rs-H0}, for the split at $\ell = 800$ instead of 1000.

We have some additional evidence of angular-scale dependence of $\omega_{\rm m}$ inferences from \citep{aylor2017}. They used a $\tau$ prior and South Pole Telescope (SPT) TT data from $650 < \ell < 3000$ to determine best-fit parameters and compared them with the parameters derived from the full Planck CMB power spectra data. They found the $\chi^2$ for the parameter differences exceeded in only 3.2\% of simulations. A major driver of this low PTE (probability to exceed) was the matter density. When both the SPT and Planck data sets were restricted to the region of sky and angular scales measured by both, the low PTE went away. As data at $\ell >1800$ is progressively added to the SPT data, increasing $\ell_{\rm max}$ to 2000, then 2500, and then 3000, the matter density decreases at each step, pulling $\omega_{\rm m}$ downward from the best-fit Planck value. Intriguingly, SPTpol data at $\ell > 1000$ also favor the same low matter density of $\omega_{\rm m} \simeq 0.13$ \cite{henning2018} while the $\ell < 1000$ data are more consistent with $\omega_{\rm m}$ from all the Planck CMB power spectra data. 

What to conclude from this scatter in inferences of $\omega_{\rm m}$ is not at all clear. The significance of the SPT trend with $\ell_{\rm max}$ is unclear. Simulations indicate that the shift in $\omega_{\rm m}$ from 1800 to 3000 is not unexpectedly large. What is potentially unusual is the absence of any scatter in the trend, but any attempt to quantify how unexpected this is would suffer from the usual problem of {\it a posteriori} statistics. Also, what are we to make of the fact that the trend is reverse to what we see from {\it Planck} where smaller angular scales deliver a larger $\omega_{\rm m}$? Perhaps this is not the influence of the radiation-driving envelope, but instead the stabilizing influence of gravitational lensing, which becomes relatively more important as a source of information about the matter density at smaller angular scales. 

We conclude that, although the situation is not yet clear, we may already be seeing, in the CMB data, additional evidence in favor of a cosmological solution to the $H_0$ discordance. Low-noise and high angular resolution measurement of temperature anisotropy over more sky than observed by SPT will potentially shed light, and may be coming soon from data already acquired by the Atacama Cosmology Telescope (ACT) collaboration. Improved measurements of polarization anisotropy at intermediate to small angular scales will help as well, as will tighter determination of the matter density from CMB lensing reconstructions. We can expect these from the SPT-3G \cite{benson2014} and AdvACT \cite{thornton2016} surveys now under way.

\section{Summary and Conclusions}

As a guide to ourselves and others, we have attempted to consider the broadest possible set of potential cosmological solutions to reconcile distance ladder, BAO and CMB observations. We divided the solutions into those that do not depart substantially from \lcdm prior to recombination, and those that do. 

Before exploring the possible cosmological solutions we reviewed the prediction of \lcdm for the Hubble constant and the comoving sound horizon, with special attention paid to the $r_{\rm s}^{\rm drag}-H_0$ plane. We saw that, with $\theta_{\rm s}^\star$ determined highly precisely, the spread of \lcdm predictions in this plane is almost entirely due to variation in the inferred matter density, $\omega_{\rm m}$. Further, we saw that variation of $\omega_{\rm m}$ does not generate movement in the plane that could simultaneously reconcile with the \citetalias{riess2019} value of $H_0$ and the relatively model-independent constraint on $H_0 r_{\rm s}^{\rm drag}$ one can infer from BOSS BAO + Pantheon supernova uncalibrated distance measurements. We also saw that this combination of data prefers a lower value of $r_{\rm s}^{\rm drag}$ than predicted by \lcdm. 

We divided the post-recombination solutions into those with a high $r_{\rm s}^\star$, as preferred when one assumes the \lcdm model, and those with a low $r_{\rm s}^\star$, as preferred by the distance ladder. For the high $r_{\rm s}^\star$ ones we discussed two ways of circumventing the distance ladder determination of $r_{\rm s}^\star$: that proposed by Raveri \cite{raveri2019}, and new-physics alterations to the supernova luminosity-flux-distance relationship to violate the usual relationship between lumniosity distance and angular-diameter distance inspired by \cite{csaki2002}. Raveri's solution relies on the introduction of a large number of additional parameters, which we find somewhat discouraging. It is possible that new photon interactions, to make distant supernovae {\em brighter} could solve the sound horizon tension. But such a solution would have to also explain the consistency of the shape of the supernova distance-redshift relation with \lcdm -- a consistency that risks becoming coincidental. 

We then discussed two post-recombination solutions with low sound horizons. The first was a very general class of potential solutions we called ``late-time confusion sowing.'' In this general class of models the CMB anisotropies are altered, and new ones are added, by effects important after recombination. The goal here is to increase the inference of the matter density and thereby lower the sound horizon. We argue that with this general class of models one cannot reduce $r_{\rm s}^\star$ sufficiently without altering $\theta_{\rm s}^{\rm EQ}$ by too much. 

Our second type of post-recombination solution with a low $r_{\rm s}^\star$ achieves, stated more precisely, a prediction of a low distance-ladder-determined $r_{\rm s}^\star.$ This predicted quantity is reduced by post-recombination evolution of the acoustic feature in the matter (or galaxy) two-point correlation function. This evolution appears to require transport of mass. We argue that the associated peculiar velocities, and the success of BAO reconstruction, pose a major challenge to a successful implementation of such a solution. 

We grouped our pre-recombination solutions into four categories: late-time confusion sowing, sound speed reduction, high-temperature recombination, and increased $H(z)$. Models in the late-time confusion sowing category (of which there are several in the literature) can not simultaneously bring both $H_0$ and $r_{\rm s}^\star$ into agreement. Models that reduce the sound horizon by reducing the sound speed are unlikely to work as we already infer the sound speed fairly directly from its influence on the zero-point of the acoustic oscillations. In analyses assuming \lcdm this sound speed inference is the chief source of information about the baryon-to-photon ratio, which is in agreement with inferences from light element abundances \cite{planck2016-l06}. 
 
High-temperature recombination is an exotic solution that would require something like time variation of the fine structure constant. Constraints in the \lcdm{+}$\alpha$ model space have been studied \cite{hart2018}. The constraints on the allowed variation of the fine structure constant are substantially tighter than the variation that would be required to reduce the sound horizon by 7\%. Although this is a specific model, we expect the result is general: the damping tail, which is highly sensitive to the history of recombination, and quite precisely measured, is not likely to allow solutions of this type. 
 
We also considered lowering the sound horizon by faster-than-adiabatic loss of energy from the photon background. Such a scenario would reduce the conformal time to recombination, and hence the sound horizon. In general, such a scenario would be tightly constrained by spectral measurements. These can, in principle, be evaded, as pointed out by \cite{chiang2018}, but, as we point out, face the same challenge from the CMB damping tail as is the case for the high-temperature recombination scenario we just discussed.

The final category is the set of solutions that introduces new components to increase $H(z)$ in the decade of scale factor evolution prior to recombination. We see these as the most likely category of solutions. They are also tightly constrained by the data. Changes here have an influence on the dynamics of mode evolution at horizon crossing, for all observable modes, and therefore have, in general, a significant influence on CMB power spectra. We discussed the radiation-driving envelope, and how its shape will in general be altered with any altering of $H(z)$ in this redshfit range. We speculated that we may already be seeing evidence of this altering of $H(z)$ in the oscillatory residuals in fits of \lcdm parameters to Planck TT data, the oscillatory residuals largely responsible for the anomalously high $A_{\rm L}$ and variations of \lcdm-based matter density inferences across angular scale. 

Models which posit new additional components include \lcdm{+}$N_{\rm eff}$ and the scalar field models explored in \cite{poulin2018,agrawal2019,lin2019a}. It is notable that none of these models alters the Planck+BAO predictions to be completely consistent with the central \citetalias{riess2019} value for $H_0$. They mostly serve to reduce the tension by broadening the uncertainty some. Perhaps further exploration of variation to $H(z)$ will provide us with a more complete resolution of the tension. 

We have been explicitly providing guidance to model builders. But our work has guidance for future observations as well. We can expect important clues, or a tightening of constraints, from a variety of CMB measurements. They include improved measurements of the CMB temperature anisotropy at $1400 < \ell < 3000$ (the angular scales for which Planck did not achieve the cosmic-variance limit), polarization measurements from $\ell \simeq 100$ to 4000, and reconstructions of the CMB lensing spectrum for precise and less radiation-driving-envelope-sensitive inferences of the matter density.

ω\section*{Acknowledgements} We would like to thank J. Chluba, F.-Y. Cyr-Racine, N. Kaloper, R. Laha L. Lancaster, L. Page, V. Poulin, and M. Raveri for useful conversations, and A. Lewis for providing some of the Monte Carlo chains used in this work. 

\appendix

\section{Visibility-averaged $\bar r_{\rm s}$ and $\bar r_{\rm d}$}
\label{app:visrsrd}

As usually defined, the sound horizon, $r_{\rm s}^\star$ or $r_{\rm s}^{\rm drag}$, and the damping scale, $r_{\rm d}^\star$, are quantities which come from performing integrals up to a certain cutoff redshift. However, the observables that we are attempting to describe with these quantities are sensitive to a range of redshifts around this cutoff. For example, the midpoint of the CMB visibility function is at around $z_\star\,{\simeq}\,1100$, but the full width at half maximum is around 200, with this entire range of redshifts contributing at some level to the way in which the sound horizon is imprinted in the final shape of the CMB power spectra. To better capture possible model changes across this range of redshifts, we introduce the ``visibility-averaged'' $\bar r_{\rm s}$ and $\bar r_{\rm d}$, defined as

\begin{align}
\bar r_{\rm s} &= \int_0^\infty dz \; g_{\rm vis}(z) \, r_{\rm s}(z) \\
e^{-(\bar r_{\rm d}/r_0)^2} &= \int_0^\infty dz \; g_{\rm vis}(z) \, e^{-(r_{\rm d}(z)/r_0)^2}
\end{align}
where the fact that the averaging happens with $r_{\rm d}(z)$ in the exponential reflects the exponential suppression of the damping, and $r_0\,{\equiv}\,30\,{\rm Mpc}$, which corresponds to $\ell\,{=}\,1500$, is where we expect most of the damping information to be coming from in the \planck data. 

This choice of derived parameters better captures the impact of model changes. As an extreme example, consider a model which looked exactly like \LCDM at $z> z_\star$, then had a significantly different $H(z)$ at $z<z_\star$. Since $r_{\rm s}^\star$ and $r_{\rm d}^\star$ involve integrals only up to $z_\star$, they would be unchanged in the modified model despite the fact that we would expect a large change to the CMB spectra. Conversely, $\bar r_{\rm s}$ and $\bar r_{\rm d}$ would capture the impact of changes anywhere in the main support of the visibility function.

Another advantage of $\bar r_{\rm s}$ and $\bar r_{\rm d}$ is that they give us a straightforward way to judge the relative sensitivity at various redshifts near last-scattering to changes in the Hubble rate (the very type of changes which we argue are likely to play a role in any cosmological resolution to the tension). To do so, we can compute a functional derivative of $\bar r_{\rm s}$ and $\bar r_{\rm d}$ with respect to $H(z)$. This is exactly the quantity plotted in Fig.~\ref{fig:integrand}. The derivation proceeds by writing out the integral for $\bar r_{\rm s}$ or $\bar r_{\rm d}$ within the expression for $\bar r_{\rm s}$ or $\bar r_{\rm d}$, e.g.
\begin{align}
\bar r_{\rm s} &= \int_{0}^{\infty} dz^\prime \; g_{\rm vis}(z^\prime) \int_{z^\prime}^\infty dz \; \frac{d r_{s}}{dz}(z),
\end{align}
and then taking a functional derivative with respect to $\ln H(z)$, repeatedly making use of the fact that
\begin{align}
\frac{\delta}{\delta \ln\! H(z)} \int_{z_\star}^{\infty} dz^\prime \frac{f(z^\prime)}{H(z^\prime)}
= \left\{
\begin{array}{cc}
    -\dfrac{f(z)}{H(z)} & z>z_\star \vspace{2mm} \\
    0 & {z<z_\star},
\end{array}\right.
\end{align}
where $f(z)$ is any other function that does not depend on the Hubble rate. After differentiating several terms and simplifying the resulting expressions, we arrive at
\begin{widetext}
\begin{align}
\frac{\delta}{\delta \ln\! H(z)} \bar r_{\rm s}
&= \frac{dr_{\rm s}^\star}{dz}(z) \big[e^{-\tau(z)} - 1\big] 
- g_{\rm vis}(z)r_{\rm s}(z) 
+ \frac{d\tau}{dz}(z) \int_z^{\infty} dz^\prime g_{\rm vis}(z^\prime)r_{\rm s}(z^\prime) \label{eq:rsvis} \\
\frac{\delta}{\delta \ln\! H(z)} e^{-(\bar r_{\rm d}/r_0)^2} 
&= \frac{dr_{\rm d}^2}{dz}(z) \big[e^{-\tau(z)} - 1\big] \big[-e^{-(r_{\rm d}(z)/r)^2)}/r^2\big] 
- g_{\rm vis}(z)e^{-(r_{\rm d}(z)/r)^2)}
+ \frac{d\tau}{dz}(z) \int_z^{\infty} dz^\prime g_{\rm vis}(z^\prime)e^{-(r_{\rm d}(z^\prime)/r)^2)} \label{eq:rdvis}
\end{align}
\end{widetext}
The curves plotted in Fig.~\ref{fig:integrand} differ from these expressions only in that one final chain rule is performed to compute $\delta \bar r_{\rm d} / \delta \ln H(z)$ and an extra factor of $z$ is included in each which corresponds to taking the functional derivation with respect to $H(\ln z)$ rather than $H(z)$.

We note that in taking this functional derivative, we have kept $x_e(z)$ (and hence $x_e$ as a function of temperature) unchanged. This would only be true if recombination happened in thermal equilibrium, which is a poor approximation. For the sound horizon calculation, $x_e(z)$ appears only inside of $g_{\rm vis}(z)$, and the final two terms in Eqn.~\ref{eq:rsvis} correspond to varying $g_{\rm vis}(z)$ with $x_e(z)$ held constant. The term we have not calculated, which would come from a chain rule variation of $x_e(z)$, can however be shown to go in the opposite direction and hence at least partially cancel. This is because more rapid expansion leads to increased number density of free electrons at a given scale factor \cite{zahn2003}, canceling the fact that with more rapid expansion we also reach a given scale factor in less time. For the case of the damping scale calculation, a factor of $x_e(z)$ additionally appears in the photon mean-free-path, which will add rather than cancel. Although we have not calculated this effect, we do not expect this to change the qualitative fact that $\bar r_{\rm d}$ is more sensitive to changes in $H(z)$ at later times as compared to $\bar r_{\rm s}$.

\bibliography{marius,Planck_bib,HubbleHunters_Lloyd,SoundOfDark}

\end{document}